\documentclass[12pt]{article}
\usepackage{graphicx}
\usepackage{float}
\usepackage[affil-it]{authblk}
\usepackage{newtxtext,newtxmath}
\usepackage[letterpaper,margin=1in]{geometry}
\usepackage{hyperref}
\hypersetup{colorlinks,
citecolor=blue
}
\usepackage{xcolor}
\usepackage{authblk}
\usepackage{mathtools}
\usepackage{amsmath}

\DeclarePairedDelimiter\ket{\lvert}{\rangle}
\DeclarePairedDelimiterX\braket[2]{\langle}{\rangle}{#1\,\delimsize\vert\,\mathopen{}#2}

\def\scititle{Interference in Quantum Mechanics}
\title{\bfseries \boldmath \scititle}
\author[1,2,*]{Urbasi Sinha}
\author[1]{Debadrita Ghosh}
\affil[1]{Raman Research Institute, Sadashivanagar, Bengaluru-560080, India}
\affil[2]{Department of Physics and Astronomy, University of Calgary, Alberta T2N 1N4, Canada}
\affil[*]{Corresponding author: usinha@rri.res.in}
\date{}

\begin{document}
\maketitle
Physicist and Nobel Laureate Richard P. Feynman once remarked, "We choose to examine a phenomenon which is impossible, \textit{absolutely} impossible, to explain in any classical way, and which has in it the heart of quantum mechanics. In reality, it contains the only mystery. We cannot make the mystery go away by ``explaining” how it works. We will just tell you how it works. In telling you how it works, we will have told you about the basic peculiarities of all quantum mechanics" \cite{feynman1963_1965}.
The phenomenon of interference is ubiquitous in the quantum world and indeed holds within itself the explanation for many counterintuitive quantum phenomena. In this review, we choose to focus on a few ramifications and manifestations of quantum interference that have deep implications for the foundations of quantum mechanics. These include single-photon or second-order interference, two-photon or fourth-order interference and higher-order interference. 

\section{One-photon interference}
\subsection{Single-photon-sources}
Before exploring quantum interference with photons, it is important to understand what a single photon is. A photon is the smallest unit of light, representing an elementary excitation of a single mode of the quantized electromagnetic field. Each mode, labeled by its frequency ($\nu_k$), corresponds to photons with energy ($h\nu_k$), where $h$ is Planck’s constant. A single photon in the ideal state is represented as $\ket{1}_k$, where $k$ defines the mode of the field. This mode may describe the photon’s spatial distribution, spectral characteristics, or temporal structure.\\
Photons interact weakly with their environment and can travel long distances at the speed of light, resulting in minimal noise and loss. They can also be easily manipulated using linear optics. These properties make photons an excellent platform for encoding information as photonic qubits, where the quantum state of the photon—such as its polarization, momentum, or energy—serves as the carrier of information. As a result, single photons are essential for applications in quantum computation and quantum information science, enabling secure communication, precise manipulation, and efficient data processing.\\
An ideal single-photon source produces one photon at a time in a specific mode, ensuring that the photon’s properties remain consistent with every emission. In practice, a perfect single-photon source cannot be achieved due to losses and the presence of multiphoton emissions. In terms of statistics, Poissonian and super-Poissonian distributions can be explained by classical wave theory, whereas sub-Poissonian statistics cannot. The observation of sub-Poissonian statistics provides strong evidence for the quantized nature of light.\\
There are various physical methods to generate single photons. Single-photon sources can be broadly classified into two categories: deterministic and probabilistic. Deterministic sources rely on systems such as colour centres, quantum dots, and single atoms, producing photons at fixed intervals. Probabilistic single-photon sources often utilize phenomena like Spontaneous Parametric Down-Conversion (SPDC) in nonlinear crystals, where photon pairs are generated by a pump laser. A key advantage of this method is that one photon (called the idler or heralding photon) signals the creation of the other photon (called the signal or heralded photon). In principle, the exact timing of the signal cannot be predicted. However, by using a pulsed pump laser, the heralding signal can be predicted within the pulse width time. For a more detailed discussion on different types of single-photon sources and their properties, readers are encouraged to consult the references \cite{Sinha:19, SINHA20231}.
\subsection{Interference in Quantum Mechanics}
Quantum interference is rooted in the concept of the superposition of probability amplitudes from processes contributing to a given phenomenon. As Dirac famously stated, "Each photon interferes only with itself." He explained that in quantum mechanics, the entities that interfere are not particles but probability amplitudes for specific events. It is the fact that these probability amplitudes add like complex numbers that give rise to all quantum mechanical interferences \cite{Dirac1958-ke}.\\ In classical mechanics, superposition produces a distinct and definite new state. In contrast, in quantum mechanics, superposition does not result in a definite new state but instead alters the probabilities of the system being observed in its basis states. In classical mechanics, interference is described through the division of amplitude, which inherently leads to a division of energy. In quantum mechanics, however, interference is described through the division of the wave function, which does not imply a division of energy, as a photon cannot be divided. In the case of a Mach-Zehnder interferometer, two physical waves with independent energies can be mutually coherent, allowing them to interfere. In the quantum picture, however, each individual photon simultaneously exists in both arms of the interferometer, with finite probability amplitudes that interfere with one another.\\
In classical wave theory, when two coherent waves interfere, their amplitudes add up algebraically. The waves may come from two different sources, such as two slits in a double-slit experiment or two arms in a Mach-Zehnder interferometer. Let’s consider two waves travelling in the same direction. The electric fields of these waves at a given point are $E_1$, and $E_2$, and we assume they are coherent. The total electric field $E_{total}$ is the sum of the individual electric fields. For simplicity, let’s assume the two waves have the same amplitude $A$, so: $E_1=A \cos(\omega t + \phi_1)$, and $E_2=A \cos(\omega t + \phi_2)$. The total electric field will be: $E_{total}=A \cos(\omega t+\phi_1)+A\cos(\omega t+\phi_2)=2A \cos(\frac{\phi_1+\phi_2}{2})\cos (\frac{\phi_1-\phi_2}{2})$. The intensity is proportional to the square of the total amplitude: 
\begin{equation}
    \label{g2_eq1}
    I_{total}\propto\vert E_{total}\vert^2 =4 A^2 \cos^2(\frac{\phi_1+\phi_2}{2})\cos^2(\frac{\phi_1-\phi_2}{2})
\end{equation}
$\cos(\frac{\phi_1+\phi_2}{2})$ describes a modulation factor that accounts for the combined effect of both phases. The factor $\cos^2(\frac{\phi_1-\phi_2}{2})$ reflects how the interference varies as a function of the phase difference, $(\phi_1-\phi_2)$ and this causes the intensity to oscillate between constructive interference and destructive interference. The amplitude of the resulting wave is directly related to the sum of the amplitudes of the individual waves. The electric fields $E_1$
and $E_2$ are added (or subtracted), meaning they divide the total amplitude into contributions from each wave.\\
In quantum mechanics, a system is described by a wave function $\psi$. The wave function $\psi(x)$ at a position $x$ gives the probability amplitude for the photon to be found at that location. The probability of detecting the photon at $x$
is proportional to the square of the magnitude of the wave function: $P(x)=\vert \psi(x)\vert^2$. This gives us a probability distribution, which tells us where the photon is most likely to be detected, but it is not a deterministic prediction. The photon does not "choose" a single path or location until it is observed; rather, it exists in a superposition of all possible paths. In a Mach-Zehnder interferometer, the photon takes both paths simultaneously, and the two paths are described by separate probability amplitudes, $\psi_1$, and $\psi_2$. The total wave function at the detector is the sum of these probability amplitudes: $\psi_{total}=\psi_1+\psi_2$. The probability of detecting the photon is related to the square of the total wave function: 
\begin{equation}
    \label{g2_eq2}
    P_{det}=\vert\psi_{total}\vert^2=\vert\psi_1\vert^2+\vert\psi_2\vert^2+2 \mathcal{R} (\psi_1^*\psi_2)
\end{equation}
The photon’s energy is not divided between the two arms of the interferometer; instead, the probabilities for detecting the photon are modified due to the interference between the probability amplitudes.
\subsection{Interferometer components}
\begin{figure}[H]
\centering
\includegraphics[scale=0.55]{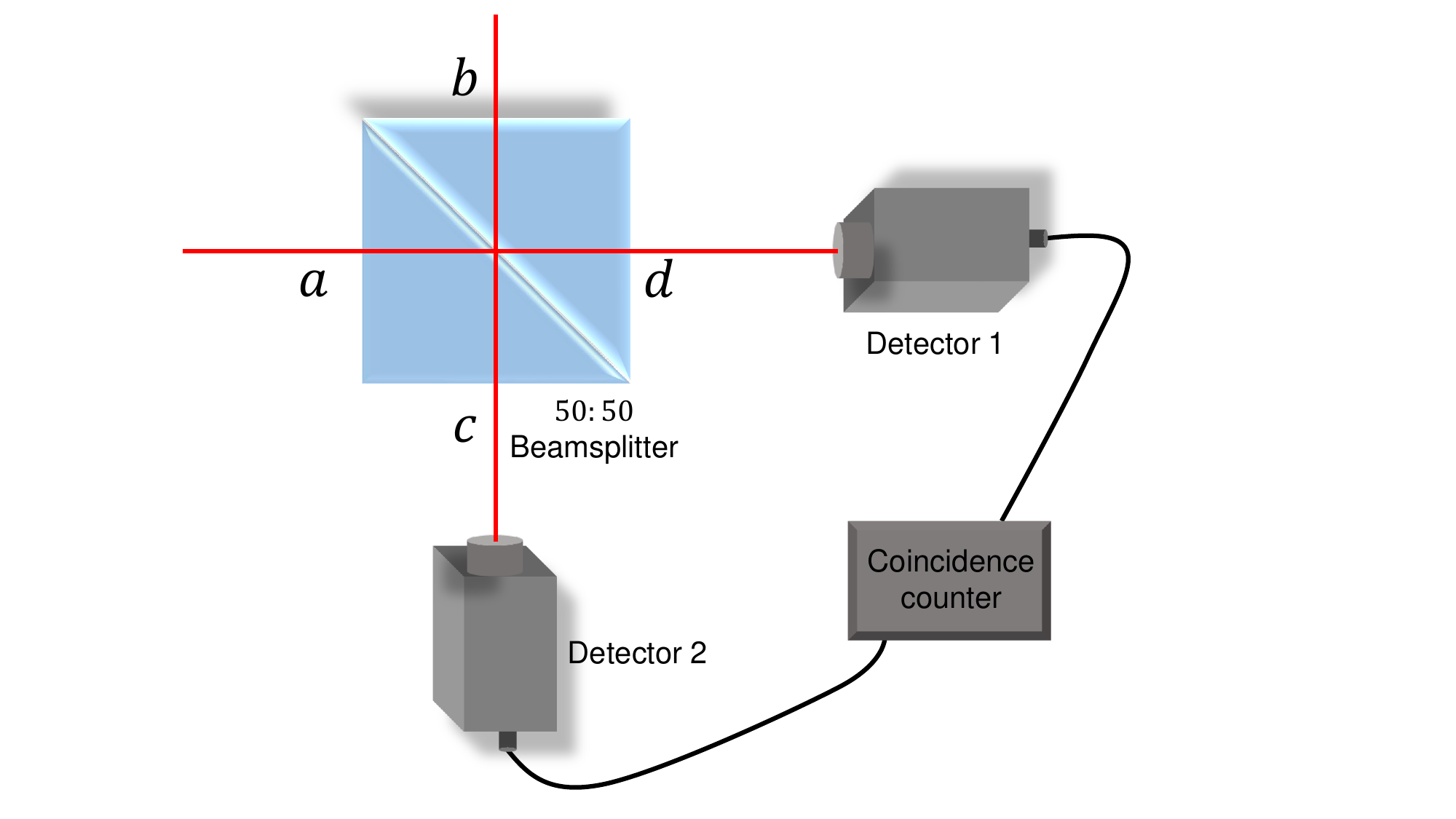}
\caption{The schematic illustrates the HBT interference setup. A $50:50$ BS is shown with two input ports ($a$ and $b$) and two output ports ($c$ and $d$). Detectors are positioned at the $c$ and $d$ output ports to measure the intensities. The correlation between the measured intensities is determined using a coincidence counter.}
\label{ge_fig}
\end{figure}

In photonic architectures, an interferometer typically consists of passive linear optical components, such as Beam-Splitters (BS), mirrors, and waveplates. It has input ports where light enters and output ports that record either photon counts in the single-photon regime or photocurrent proportional to the laser intensity. A BS is a key component in many practical interferometers. It splits incident light into two separate beams—one transmitted and the other reflected—at a specific ratio. A BS has two input ports and two output ports. Light entering through one input port is split, with the transmitted and reflected components measured at the two output ports. Alternatively, light can enter through both input ports simultaneously, and the resulting split beams can be observed at both output ports. While a BS is a straightforward device from the perspective of classical physics, its behaviour becomes complex and non-trivial in the context of quantum mechanics. The statistical behaviour of photons at the BS output changes, giving rise to fundamental quantum phenomena such as quantum superposition and randomness. A BS can also create entanglement between the output fields \cite{PhysRevA.65.032323,Makarov2021}.\\
In the quantum mechanical description, the input and output fields of a BS are represented by operators. As shown in Fig.\ref{ge_fig}, the input ports of the BS are labeled $a$ and $b$, while the output ports are labeled $c$ and $d$. A mathematical relationship connects the input and output operators. The input modes $a$ and $b$ are associated with the annihilation and creation operators $\hat{a}$, $\hat{a}^\dagger$ and $\hat{b}$, $\hat{b}^\dagger$, respectively. Similarly, the output modes $c$ and $d$ correspond to the annihilation and creation operators $\hat{c}$, $\hat{c}^\dagger$ and $\hat{d}$, $\hat{d}^\dagger$, respectively. The input-output transformation matrix can be expressed by replacing the classical electric field components with their corresponding quantum operators.
\begin{equation}
\label{g2_eq3}
\begin{pmatrix}
  \hat{c} \\
  \hat{d} \\
\end{pmatrix}
=
\begin{pmatrix}
R & T\\
T & R
\end{pmatrix} 
\begin{pmatrix}
  \hat{a} \\
  \hat{b} \\
\end{pmatrix}
\end{equation}
Here, $T$ and $R$ represent the transmission and reflection coefficients of a BS, respectively. Both $T$ and $R$ are complex numbers. For a lossless BS, the BS matrix is unitary, meaning $\vert T\vert^2+\vert R\vert^2=1$, and also $R^*T+T^*R=0$. Starting from the Eq.\ref{g2_eq3}, it can also be expressed as, 
\begin{equation}
\label{g2_eq4}
\begin{pmatrix}
  \hat{a} \\
  \hat{b} \\
\end{pmatrix}
=
\begin{pmatrix}
R^* & T^*\\
T^* & R^*
\end{pmatrix} 
\begin{pmatrix}
  \hat{c} \\
  \hat{d} \\
\end{pmatrix}
\end{equation}
In the case of a $50:50$ BS, the transmission and reflection coefficients each correspond to $50\%$ of the incident light, giving a complex amplitude of $\frac{1}{\sqrt{2}}$. If the BS is symmetric, the reflection introduces a phase shift of $\frac{\pi}{2}$ relative to the transmission. So, the BS matrix is expressed as $\frac{1}{\sqrt{2}}\begin{pmatrix}
1 & i\\
i & 1
\end{pmatrix}$.\\
A basic interferometer can be created using a BS, which demonstrates the Hanbury Brown-Twiss (HBT) effect. Although the setup is simple, the HBT interference effect remains a landmark experiment in quantum optics \cite{BROWN1956}.
\subsection{Theoretical derivation of HBT interference}
In this effect, light passes through a BS and reaches two detectors at the two output ports. The correlation of the intensities measured by these detectors is recorded. The key observation of the HBT effect is that the correlation between the light intensity at two different detectors can reveal the properties of the light. The second-order correlation function $g^{(2)}(\tau)$
quantifies the correlation between the intensities measured at the two detectors at different times. If $l_1$ is the optical path length traversed by light from the BS to the detector at output port $c$, and $l_2$ is the optical path length traversed by light from the BS to the detector at output port $d$, then the time delay $\tau$ is given by $\tau=\frac{\vert l_1-l_2\vert}{c}$,  where $c$ is the speed of light. If light with intensity $I$
is incident on a BS, $I_c(t)$ and $I_d(t+\tau)$ represent the intensities measured by the detectors at output ports $c$ and $d$ at times $t$ and $(t+\tau)$, respectively. The normalized correlation function of the light intensity is then given by:
\begin{equation}
    \label{g2_eq5}
    g^{(2)}(\tau)=\frac{\langle I_c(t)I_d(t+\tau) \rangle}{\langle I_c(t)\rangle \langle I_d(t+\tau)\rangle}
\end{equation}
$g^{(2)}$ measures how correlated the intensities at the two detectors are. For a $50:50$ beam splitter, when $\tau=0$, the $g^{(2)}$ equals 1 for classical light.\\
In quantum mechanics, photons are described using the Fock state formalism. When a single photon is incident on the BS, its state is $\ket{1}_a \ket{0}_b$. A single photon in a specific mode is represented by applying the creation operator to the vacuum state of that mode, expressed as $\ket{1}_a \ket{0}_b=\hat{a}^\dagger \ket{0}_a \ket{0}_b$. From Eq. \ref{g2_eq4}, $\hat{a}^\dagger=R\hat{c}^\dagger+T\hat{d}^\dagger$. Therefore, the output state is,
\begin{equation}
    \label{g2_eq6}
    \ket{\psi_{HBT}}=(R\hat{c}^\dagger+T\hat{d}^\dagger)\,\ket{0}_c \ket{0}_d = R\,\ket{1}_c \ket{0}_d + T\,\ket{0}_c \ket{1}_d
\end{equation}
This represents the superposition state of a photon being in output port $c$ or a photon being in port $d$. For a $50:50$ BS, the output state is,
\begin{equation}
    \label{g2_eq7}
    \ket{\psi_{HBT}}=\frac{1}{\sqrt{2}}(i\ket{1}_c \ket{0}_d + \ket{0}_c \ket{1}_d)
\end{equation}
The average number of photons in port $c$ is given by the expectation value of the number operator $\hat{a}^\dagger\hat{a}$, which evaluates to $\langle \psi_{HBT}\vert \hat{a}^\dagger\hat{a}\vert\psi_{HBT}\rangle=\vert R \vert^2=\frac{1}{2}$. The same result holds for output port $d$.\\
If we repeat the measurement many times, half of the incident photons are detected by the detector in port $c$, and the other half are detected by the detector in port $d$. This result aligns perfectly with classical predictions. However, when examining the correlation between the detected photons in both output ports, non-classicality becomes evident. In calculating $g^{(2)}$ using Eq. \ref{g2_eq5} the average intensity $\langle I_c(t)I_d(t+\tau) \rangle$ is replaced by photon number operator. The expectation value of the photon number operator then gives the photon count in practice. At $\tau=0$, the expression for $g^{(2)}$ is then given by,
\begin{equation}
\label{g2_eq8}
g^{(2)}(\tau=0)=\frac{\langle \hat{c}^\dagger\,\hat{c}\,\hat{d^\dagger}\,\hat{d}\rangle}{\langle\hat{c^\dagger}\,\hat{c}\rangle\,\langle \hat{d^\dagger\,\hat{d}}\rangle}
\end{equation}
This result yields a value of $0$, which is a hallmark of quantum behaviour. This outcome serves as a crucial test for the single-photon source, confirming that it indeed generates photons one at a time.
\newpage
\section{Two-photon interference}
The interference of two photons generated simultaneously through parametric down-conversion was analyzed in 1986 \cite{PhysRevA.34.3962}. The study explored fourth-order interference effects, distinct from standard second-order interference involving single-field quantities. These effects were observed by measuring the joint probability of detecting both photons at two points in the interference plane using photoelectric detectors. The joint detection probability exhibited cosine-like modulation with visibility approaching $100\%$ under ideal conditions.
In 1987, C.K. Hong, Z.Y. Ou, and L. Mandel conducted a seminal experiment demonstrating a quantum interference effect when indistinguishable photons were incident on a $50:50$ BS \cite{PhysRevLett.59.2044}. This phenomenon, known as "photon bunching," resulted in the two photons exiting the BS together in the same output port. The authors have demonstrated the measurement of extremely short time intervals between two photons, providing insights into the length of the photon wave packet.
This was further investigated by \cite{PhysRevLett.59.1903}, where the researchers demonstrated that, unlike second-order interference, which is absent due to the lack of a definite phase relationship between photons, fourth-order interference manifested through joint probability measurements of photon detection at two positions. The interference exhibited $100\%$ visibility, revealing nonclassical features. These results aligned with quantum predictions and contrasted with classical wave theories, which predicted a maximum visibility of only $50\%$.
The theory of fourth-order interference using a BS has been explored in detail \cite{PhysRevA.37.1607}, including the conditions under which classical and quantum fields exhibit interference. 
In 1988, Y.H. Shih, and C.O. Alley conducted an experiment \cite{PhysRevLett.61.2921} with correlated photons generated from a nonlinear crystal via down conversion. They measured both coincidence counts and single counts and then calculated the ratio between them. Polarizers were placed in front of the detectors to create different polarization eigenstates. The results showed that the ratio reached its maximum when the polarization axes were perpendicular and its minimum when the axes were parallel.\\
Around the same period, Fearn and Loudon explored theoretical frameworks of two-photon interference \cite{Fearn:89} for different types of photon sources. The authors calculated the second factorial moments of photocounts and the cross-correlation function for the two output arms of a lossless BS. When photon pairs enter the BS through separate input arms and arrive simultaneously, the output state where one photon is detected in each arm is strongly suppressed. This suppression is evident in the photo count cross-correlation, which can vanish for certain input parameters from two-atom or parametric oscillator sources and is reduced to half for light from cascade emission sources. J.G. Rarity and P.R. Tapster \cite{PhysRevA.41.5139} delved into the application of two-photon interference in the context of the Bell test experiment. Their work extended the fundamental principles established by Hong, Ou, and Mandel, focusing on utilizing two-photon interference for testing quantum nonlocality.\\
\begin{figure}
\setlength{\abovecaptionskip}{10pt} 
\centering
\includegraphics[scale=0.15]{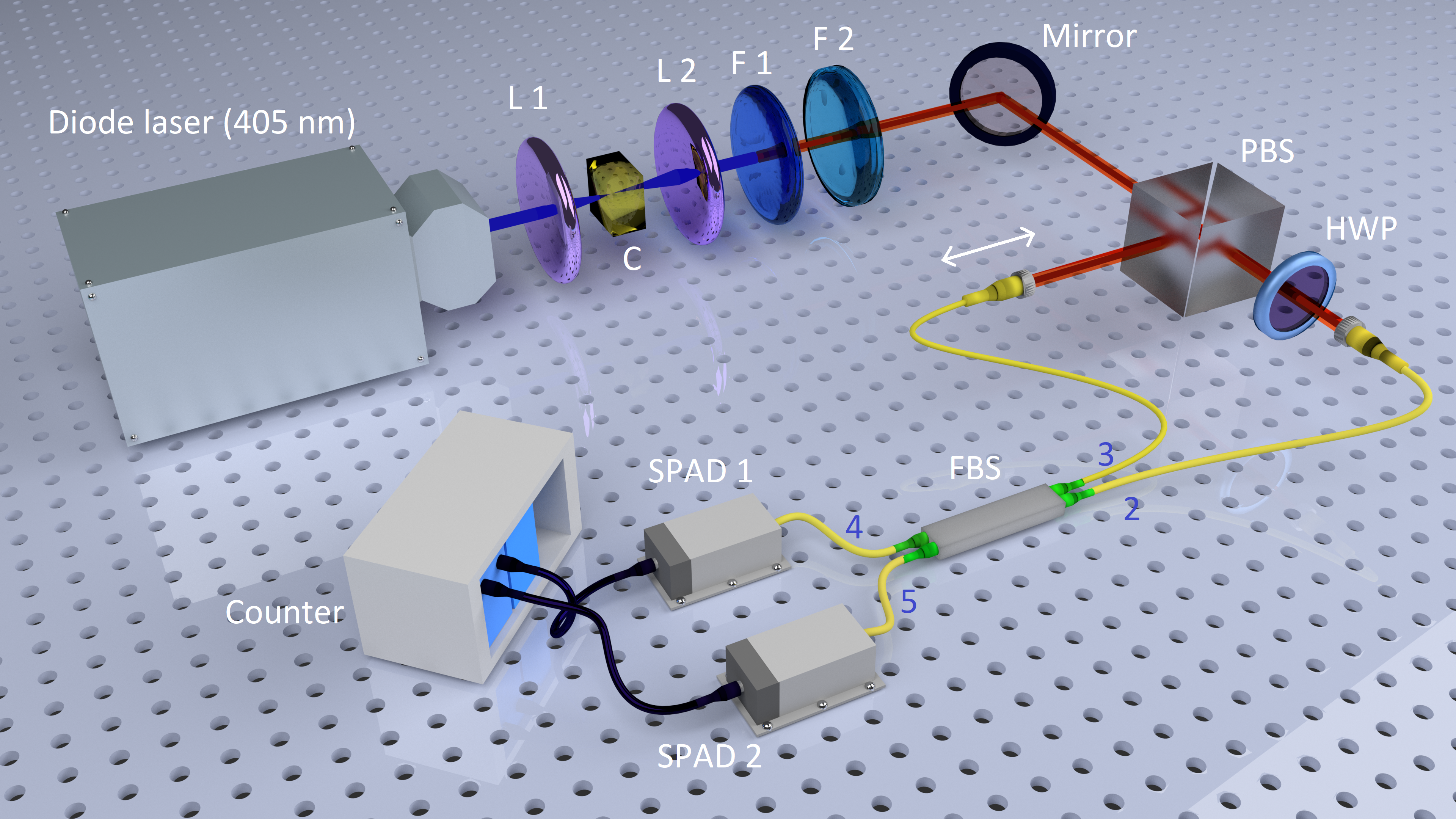}
\caption{Illustration of the HOM experimental setup, adapted from \cite{PhysRevA.100.013839}. A diode laser generates a coherent pump beam at 405 nm wavelength, which falls on a type II BBO crystal (C). A pair of lenses (L1) and (L2) is used to focus the pump beam at the crystal and to collimate the beam, respectively. The polarization of the pump beam is maintained as horizontal (H). The crystal is tilted such that the collinear phase matching condition is satisfied. In this configuration, down-conversion creates pairs of orthogonally polarized, frequency-degenerate photons at 810 nm wavelength. A long-pass filter (F1) blocks the pump beam (405 nm) and passes only the single-photon pairs (810 nm). A band-pass filter (F2) with a 3.1 nm bandwidth centred on 810 nm restricts the bandwidth of the transmitted photons and minimize any distinguishability in spectral degree of freedom. Two
photons with orthogonal polarization, in any pair, are split in two directions by a polarizing beam
splitter (PBS). Again, a half-wave plate (HWP) in one of the output arms of the PBS is used to
make both photons the same polarized. Two fiber-couplers collect photons from the same pair in two single-mode fibers; in order to maintain indistinguishability in spatial degree of freedom, and inject
them in a $2\times 2$ polarization maintaining, fused fiber beam-splitter (FBS). One of the couplers is
mounted on a motorized stage, which translates along the direction of the beam, to set a variable
delay $\delta\tau$ between the two photons. After passing through the FBS, photons are detected in
two single-photon avalanche detectors (SPAD), which are connected to a
single-photon counting module; in order to perform coincidence measurement.}
\label{fig1}
\end{figure}\\
Two-photon interference, commonly referred to as the Hong-Ou-Mandel (HOM) effect, is a fundamental phenomenon in quantum optics. Fig.\ref{fig1} shows a schematic of the experimental setup for the HOM experiment \cite{PhysRevA.100.013839}. 
The HOM effect occurs when two indistinguishable photons enter a BS from two different input ports (ports $2$, and $3$ in Fig.\ref{fig1}), leading to interference that results in the photons being detected at the same output port (either port $4$ or $5$ in Fig.\ref{fig1}), even though classical physics would predict otherwise. This two-photon interference with a balanced BS can be derived using quantum mechanical formalism.
\subsection{Theoretical derivation for HOM interference}
Consider,two optical input modes ($a$,$b$), and two output modes ($c$,$d$) of a lossless $50-50$ BS. $(a,a^\dagger)$, $(b,b^\dagger)$, $(c,c^\dagger)$, and $(d,d^\dagger)$ are the annihilation and creation operators in the BS modes, such that,
\begin{equation*}
    \ket{1;j}_a \ket{1;k}_b=a_j^\dagger b_k^\dagger \ket{0}_a \ket{0}_b,
\end{equation*} where $j$, and $k$ are photon's properties which determine how distinguishable they are. The transformation of a state during its interference on a BS can be described using a unitary operator. This operator governs the evolution of the creation operators as
$a^\dagger \rightarrow \frac{c_j^\dagger+d_j^\dagger}{\sqrt{2}}$ and $b^\dagger \rightarrow \frac{c_k^\dagger-d_k^\dagger}{\sqrt{2}}$. The output state is,
\begin{equation} 
\label{eq1}
\begin{split}
\ket{\psi_{HOM}} & = \frac{c_j^\dagger+d_j^\dagger}{\sqrt{2}}\,\, \frac{c_k^\dagger-d_k^\dagger}{\sqrt{2}} \ket{0}_c \,\ket{0}_d \\
 & = \frac{1}{2}\left(c_j^\dagger c_k^\dagger-c_j^\dagger d_k^\dagger+ d_j^\dagger c_k^\dagger-d_j^\dagger d_k^\dagger\right)\,\, \ket{0}_c\, \ket{0}_d 
\end{split}
\end{equation}
If all the properties of the photon pair are identical ($j=k$), then the second and fourth terms in Eq.\ref{eq1} cancel out. Therefore, the output state is,
\begin{equation}
\label{eq2}
    \ket{\psi_{HOM}}=\frac{1}{\sqrt{2}}\left( \ket{2}_c\,\ket{0}_d - \ket{0}_c\,\ket{2}_d\right)
\end{equation} The state is still normalized because $a^\dagger \,\ket{n}=\sqrt{n+1}\,\ket{n+1}$.\\
If the two photons are completely indistinguishable, the probability of detecting a coincidence at the output ports of the BS is $0$. However, if the photons are distinguishable, the second and third terms in Eq.\ref{eq1} do not cancel out. As a result, the output state is given by,
\begin{equation}
\label{eq3}
    \ket{\psi_{dis}}=\frac{1}{2}\,\left(\ket{2}_c\,\ket{0}_d-\ket{1}_c\,\ket{1}_d+\ket{1}_d\,\ket{1}_c-\ket{0}_c\,\ket{2}_d\right)
\end{equation}
The probability of coincidence is $\lvert\frac{1}{2}\rvert^2+\lvert-\frac{1}{2}\rvert^2=\frac{1}{2}$.\\
As the time delay $(\tau)$ between the photons is varied, the coincidence detection rate shows a dip at zero delay, which is a signature of the HOM effect. The depth of this dip serves as a key indicator of the quantum nature of the light source and the degree of indistinguishability between the photon pair. This measure is referred to as the visibility of the interference.\\
Based on the discussion above, the probability of coincidence at the output ports of the BS is $\frac{1}{2}$ when $j\neq k$ and $0$ when $j=k$, the visibility of the HOM dip can be expressed as,
\begin{equation}
\label{eq4}
    V_{HOM}=\frac{p(\tau=\infty)-p(\tau=0)}{p(\tau=\infty)+p(\tau=0)}
\end{equation}, where $p(\tau)$ represents coincidence probability, given by,
\begin{equation}
\label{eq5}
    p(\tau)=\frac{1}{2}[1-\delta(\tau)],
\end{equation} where $\delta(\tau)$ is 
defined as $\delta(\tau)=1$ at $\tau=0$ and $\delta(\tau)=0$ otherwise.\\
In this context, it is important to note that practical simultaneous detection is limited by a finite time window. If two signals arrive within this defined interval, they are regarded as arriving simultaneously. This interval is known as the coincidence window.
\begin{figure}[H]
\centering
\includegraphics[scale=0.6]{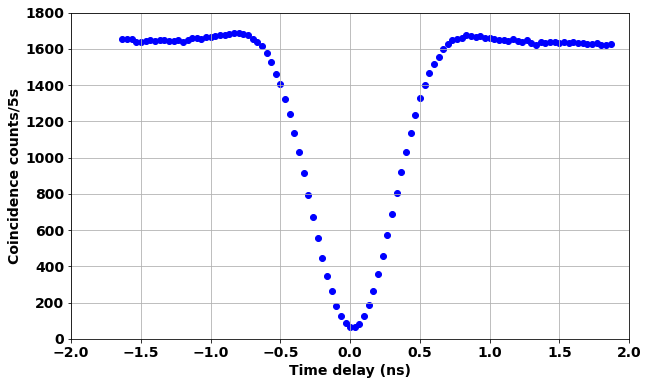}
\caption{An example of a HOM dip, where the coincidence counts are plotted as a function of the path delay between two photons, is achieved in the experimental setup presented in Fig \ref{fig1} \cite{PhysRevA.100.013839}. The data acquisition time is $5$ seconds, and for each time delay the measurement is repeated 100 times to improve averaging. Blue points represent the experimental mean. }
\label{fig2}
\end{figure}
Here, the indistinguishability between the pair of photons encompasses their spectral, polarization, and temporal overlap. If the arrival times of the two photons at the $50:50$ BS differ, a temporal delay $\tau$ is introduced between them. This delay causes the temporal mode functions of the two photons to no longer perfectly overlap. In the operator formalism, the creation operators acquire temporal dependence: $a_j^\dagger \rightarrow a_j^\dagger (t)$ and $b^\dagger_k \rightarrow b^\dagger_k(t+\tau)$. Using thes, the output state becomes
\begin{equation}
    \label{r_eq6}
 \ket{\psi_{HOM}} =   \frac{1}{2}\left(c_j^\dagger(t) c_k^\dagger (t+\tau)-c_j^\dagger (t) d_k^\dagger (t+\tau) + d_j^\dagger (t) c_k^\dagger (t+\tau) - d_j^\dagger (t) d_k^\dagger (t+\tau) \right)\,\, \ket{0}_c\, \ket{0}_d 
\end{equation}
If $\tau=0$ and $j=k$, the operators $c_j^\dagger(t)$ and $c_k^\dagger(t+\tau)$ (and similarly for the $d$ mode operators) act on the same temporal mode. In that case, the second and third terms have the same magnitude but opposite sign, leading to a commutation-like cancellation. However, if $\tau \ne 0$ even when $j=k$, the creation operators act on different temporal modes due to the delay. As a result, the cancellation of the second and third terms is no longer perfect. To quantify the partial cancellation, the temporal mode functions $\phi_j(t)$ and $\phi_k(t+\tau)$ are introduced. These functions describe the temporal profile of the photon wavepackets at times $t$ and $t+\tau$, respectively, and satisfy the normalization condition: $\int \vert \phi_j(t)\vert^2 dt=1$. For simplicity, these functions can be assumed to have a Gaussian profile. The operators act on these envelopes of wavepackets as: $a_j^\dagger(t)=\phi_j(t)\,a_j^\dagger$ and $b_k^\dagger(t+\tau)=\phi_k(t+\tau)\,b_k^\dagger$. The degree of cancellation between the second and third terms in Eq.\ref{r_eq6} depends on the overlap of the temporal mode functions: $\langle \phi_j \vert \phi_k \rangle = \int \phi_j^*(t)\,\phi_k(t+\tau) dt$. At $\tau=0$, $\phi_j(t)=\phi_k(t)$, so the overlap is $1$, leading to perfect cancellation of the second and third terms in Eq.\ref{r_eq6}. At $\tau \ne 0$, the overlap decreases. Consequently, the coincidence probability in Eq.\ref{eq5} is modified to: 
\begin{equation}
\label{eq7}
   p(\tau)=\frac{1}{2}[ 1- \vert \langle \phi_j \vert \phi_k \rangle \vert ^2] 
\end{equation}.

\subsection{Spectral distingushability}
In practice, the incident photons are not perfectly monochromatic, meaning the two photons are not spectrally identical and possess a spectral width $\sigma_\omega$. As a result, Eq.\ref{eq5} is modified by,
\begin{equation}
    \label{eq6}
    p(\tau)=\frac{1}{2}[1-e^{\sigma_\omega^2\,\tau^2}]
\end{equation}
The spectral amplitudes of the two photons can be correlated, as observed in parametric down-conversion processes. This spectral correlation is described by the joint spectral amplitude $f(\omega_1,\omega_2)$, where $\omega_1$, and $\omega_2$ represent the central frequencies of the two incident photons. The coincidence probability in this scenario has been both theoretically and experimentally demonstrated, as shown in \cite{Jin:18}, and is given by,
\begin{equation}
    \label{eq7}
    p(\tau)=\frac{1}{2}\,\left(1-\int_{0}^{\infty}\int_{0}^{\infty}d\omega_1\,d\omega_2\,\lvert f(\omega_1,\omega_2)\rvert^2\,e^{i(\omega_1-\omega_2)\tau}\right)
\end{equation}
The spectral function, $f(\omega_1,\omega_2)$ can be modified by applying a bandpass filter. Typically, the filter function can be approximated by a sinc function, which results in a bump in the wings of the HOM dip as shown in Fig.\ref{fig2}.\\
Spectral and temporal distinguishability in multi-photon interference has been explored both theoretically and experimentally, and HOM dips have been observed with different degrees of visibility, depending on the level of indistinguishability achieved in different experimental scenarios, in the ref.\cite{PhysRevA.56.1627,PhysRevA.74.063808,PhysRevA.83.062111,PMID:24022582,doi:10.1073/pnas.1206910110,Tichy_2014,LI2023109039,doi:10.1126/sciadv.adm7565,Mann:24,PhysRevLett.93.070503,PhysRevA.108.023713,Duan:24,Li:25,PhysRevLett.89.213601,Thiel:20}.\\
Furthermore, the variation in the visibility with the pump power in parametric down-conversion has been studied both theoretically and experimentally for type I and type II phase-matching crystals in \cite{PhysRevA.77.053822}.
\subsection{Application of HOM interference}
\subsubsection{Manifestation of quantumness}
A $50\%$ visibility in the HOM dip is generally regarded as the threshold distinguishing classical from quantum behavior of light. When two classical beams with equal intensities and randomly varying relative phase are incident on a 50:50 BS, the maximum achievable HOM dip visibility is limited to $50\%$ \cite{Rarity_2005}. However, in \cite{PhysRevA.100.013839}, an experiment demonstrated that nearly $100\%$ visibility in the HOM dip can be achieved even with classical fields. In the semiclassical framework of photo-detection theory, the coincidence probability is proportional to the cross-correlation of the integrated intensities at the detectors. The normalized correlation function $C(\tau)$ is derived, and it is shown that the visibility of the coincidence dip, denoted as $V$, depends on the fluctuation probability distribution of the relative phase $\phi$ between the input pulses. To further elaborate, the phase fluctuations between the input fields are critical in determining the visibility of the HOM dip. In the context of this experiment, the phase is treated as a random variable, and its fluctuation is captured by a fixed probability distribution $P(\phi)$. The probability distribution is chosen such that the average fluctuation of the phase, when weighted by $P(\phi)$, does not introduce a bias, which requires $\int P(\phi) \cos \phi\,\,d\phi=0$. This ensures that the cosine function's average value over all phase values is zero. This condition effectively represents a uniform distribution of phase fluctuations without a net phase shift. Therefore, $V$ is expressed as,
\begin{equation}
    \label{our_paper_eq1}
    V=\int P(\phi) \cos^2 \phi\,\, d\phi
\end{equation}
When the relative phase $\phi$ is uniformly randomized over $[0,2\pi]$, Eq.\ref{our_paper_eq1} shows that the $V$ reaches the classical limit of $50\%$.
However, with controlled phase distributions, such as $[0,\pi]$, $V$ can reach $100\%$, even with classical pulses. This demonstrates that the classical upper limit of $50\%$ arises from a lack of phase control rather than being an inherent boundary of classical electromagnetic field theory.\\
The experiment employs an electrical system where phase control is directly achieved through an arbitrary waveform generator. Two Gaussian amplitude-modulated sine waves, generated by the waveform generator, are input to a balanced BS (power splitter) with carrier frequencies and relative phases controlled to ensure identical inputs. The system's components, including mixers, oscilloscopes, and power splitters, are calibrated for optimal performance. The time delay between the two input signals is varied systematically, and the cross-correlation of the outputs is analyzed to determine visibility. With phase values restricted to $[0,\pi]$, with equal probability, the experiment achieves a visibility of $99.63\%$.\\
To further distinguish between quantum and classical behavior, a complementarity test was conducted by introducing an additional BS into the setup, effectively constructing a Mach-Zehnder interferometer with a phase shifter. The experiment was carried out in two scenarios: In Case I, both input arms of the second BS were open, while in Case II, one of the input arms was blocked. In Case I, both arms of the interferometer are unblocked, and the second BS recombines the photon paths. In the classical scenario, the output from the first BS enters both arms, resulting in an interference pattern with a $100\%$ correlation. In the quantum scenario, the photons are in a superposition state, and the system behaves similarly, also showing a $100\%$ coincidence. In Case II, in the classical scenario, when one arm of the interferometer is blocked, it effectively blocks half of the signal. The unblocked arm is then split into two, resulting in a $50\%$ cross-correlation compared to Case I, as the beam is only partially recombined. In the quantum scenario, when one arm is blocked, the superposition state collapses to one of two possibilities: $\ket{0,2}$ or $\ket{2,0}$, each with equal probabilities. After passing through the second BS, the output state is given by: $\frac{1}{2} \ket{0,2} + \frac{1}{2} \ket{2,0}+\frac{1}{\sqrt{2}}\ket{1,1}$. This results in a coincidence probability of $\frac{1}{2}$. The decrease in coincidence probability, combined with the blocking of half of the superposition states, leads to a $25\%$ coincidence rate compared to Case I.\\
In \cite{Fabre2022}, the quantum and classical aspects of the HOM experiment were analyzed, presenting a general expression for intensity correlations that clearly highlights the differences between a classical HOM-like dip and a quantum one.
\subsubsection{Indicator of indistinguishability}
The visibility of a HOM dip is a critical measure of the indistinguishability of input photons. Two photons are considered indistinguishable if they share the same electromagnetic field mode, meaning they have identical polarization, frequency, arrival time, and spatial profile. When photons are perfectly indistinguishable, the HOM visibility approaches $1$. This phenomenon is commonly used to assess the mode indistinguishability of photons generated through SPDC \cite{PhysRevLett.59.2044,PhysRevLett.61.2921,PhysRevA.41.5139}. However, HOM interference is not limited to SPDC-generated photons. It can also be applied to single photons produced by other physical systems, such as quantum dots and vacancy centers, to evaluate their indistinguishability \cite{doi:10.1021/nl503081n,10.1038/nnano,PhysRevLett.126.063602}.
\begin{figure}[H]
\centering
\includegraphics[scale=0.45]{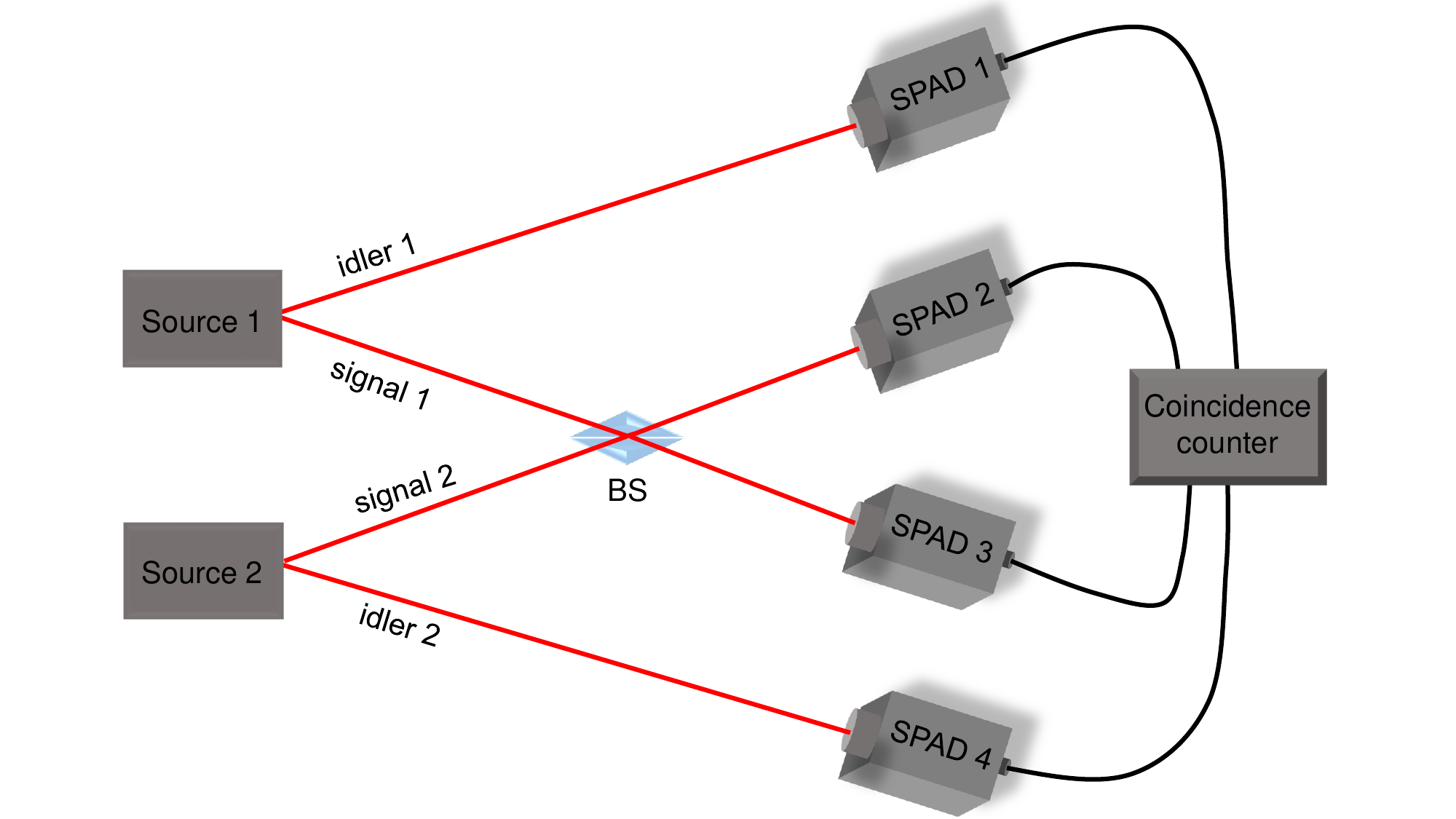}
\caption{The schematic illustrates the experimental setup for HOM interference between two photons originating from two distinct single-photon sources (source 1 and source 2). Signal photons (s1 and s2) from each source interfere at the BS, while the corresponding idler photons (i1 and i2) are used for heralding. Four single-photon detectors (SPADs) are employed to measure photon counts, and a coincidence counter records the four-fold coincidence events. }
\label{fig3}
\end{figure}
HOM interference can also occur between two photons originating from independent sources—one from source $1$ and the other from source $2$. This type of interference is crucial for quantum communication protocols, such as quantum teleportation. Fig.\ref{fig3} illustrates a general schematic of the experimental setup. Here, signal photons (s1 and s2) from both sources are directed to the two input ports of a BS, while the corresponding idler photons (i1 and i2) are used for heralding. In this scenario, Eq.\ref{eq7} is modified to reflect the four-fold coincidence probability \cite{Jin:15}.\\The four-fold coincidence, $p_4(\tau)$
can be expressed as a function of the time delay $\tau$,
\begin{multline}
  \label{eq8}
    p_4(\tau)=\frac{1}{4}\int_{0}^{\infty}\int_{0}^{\infty}\int_{0}^{\infty}\int_{0}^{\infty}d\omega_{s1}\,d\omega_{i1}\,d\omega_{s2}\,d\omega_{i2}\,\lvert f_1(\omega_{s1},\omega_{i1})\,f_2(\omega_{s2},\omega_{i2})-\\f_1(\omega_{s2},\omega_{i1})\,f_2(\omega_{s1},\omega_{i2})\,e^{-i(\omega_{s2}-\omega_{s1})\tau}\rvert^2  
\end{multline}
For $100\%$ visibility, at $\tau=0$, $p_4(\tau)$ must be $0$. As demonstrated by Eq. \ref{eq8}, this condition is satisfied if, $f_1(\omega_{s1},\omega_{i1})\,f_2(\omega_{s2},\omega_{i2})=f_1(\omega_{s2},\omega_{i1})\,f_2(\omega_{s1},\omega_{i2})$. This condition is met when the two joint spectral amplitudes are identical.
\begin{equation}
    \label{eq9}
f_1(\omega_{s1},\omega_{i1})=f_2(\omega_{s2},\omega_{i2})
\end{equation}
Additionally, both functions must be separable such that they can be expressed as a product of independent components.
\begin{equation}
    \label{eq10}
f(\omega_s,\omega_i)=\mathcal{F}_s(\omega_s)\mathcal{F}_i(\omega_i)
\end{equation}
Eq.\ref{eq9} indicates that the two sources emit indistinguishable photons and Eq.\ref{eq10} indicates that the sources are spectrally pure, meaning that the photons emitted by the sources have well-defined spectral properties, such as a narrow bandwidth and a single frequency component. The integral in Eq.\ref{eq8} is evaluated only for the signal photons, as they contribute to the HOM interference. Eq.\ref{eq8} exclusively for the signal photons,
\begin{equation}
    \label{eq11}
    p_4(\tau)=\frac{1}{4}\int_{0}^{\infty}\int_{0}^{\infty}d\omega_{s1}\,d\omega_{s2}\,\lvert \mathcal{F}_{s1}(\omega_{s1})\mathcal{F}_{s2}(\omega_{s2})- \mathcal{F}_{s1}(\omega_{s2})\mathcal{F}_{s2}(\omega_{s1}) e^{-i(\omega_{s2}-\omega_{s1})\tau}\rvert^2
\end{equation}
The photon spectra are considered as normalized Gaussian functions ($\mathcal{F}_s(\omega_s)=\sqrt{\frac{2}{\pi\,\sigma_s^2}}\exp{[-(\frac{\omega_s-\omega_{s0}}{\sigma_s})^2]}$). After integration Eq.\ref{eq11} simplifies to,
\begin{equation}
    \label{eq12}
    p_4(\tau)=\frac{1}{2}-\frac{\sigma_{s1}\,\sigma_{s2}}{\sigma_{s1}^2+\sigma_{s2}^2}\,exp\left[-\left(\frac{\sigma_{s1}^2\,\sigma_{s2}^2\,\tau^2+4(\omega_{s20}-\omega_{s10})^2}{2(\sigma_{s1}^2+\sigma_{s2}^2)}\right)\right],
\end{equation}
where $\omega_{s10}$ and $\omega_{s20}$ represent the central frequencies of the photons, while $\sigma_{s1}$ and $\sigma_{s2}$ denote their respective bandwidths. Eq. \ref{eq12} shows that as the central frequencies of the photons deviate from each other, the visibility of the HOM dip decreases. Similarly, if the bandwidth of one photon becomes larger than that of the other, the visibility also diminishes.\\
HOM interference experiments between two photons generated from independent sources have been demonstrated, including experiments with two photons generated by two separate SPDC sources, as well as experiments involving one single photon generated by SPDC and one from a weak coherent pulse. In this setup, one of two configurations is typically used to generate the pump wavelength required for SPDC. In the first configuration, a mode-locked titanium-sapphire laser serves as the master laser. The laser output undergoes frequency doubling to produce the desired pump wavelength, which is then split by a BS. The reflected beam pumps one nonlinear crystal, while the transmitted beam pumps the second crystal, effectively creating two independent SPDC sources \cite{PhysRevA.67.022301}. Using this setup, HOM visibility as high as $94\%$ has been achieved, demonstrating the high level of indistinguishability between the two independent sources \cite{PhysRevLett.100.133601}.
Alternatively, the two nonlinear crystals can be pumped sequentially by directing the beam through both crystals \cite{wang}.
These configurations ensure that both crystals are pumped by the same master laser, maintaining coherence between the two-photon sources.
Alternatively, in the second configuration, two independent titanium-sapphire lasers are used to pump the two SPDC sources. To ensure synchronization between the laser pulses, an electronic feedback mechanism is employed \cite{PhysRevLett.96.240502}.\\
If one of the sources is a weak, coherent pulse, the setup is slightly modified. In this case, a BS is placed in the path of the titanium-sapphire laser beam. One of the split beams serves as the weak coherent pulse source after being attenuated by a filter, while the other beam is directed towards the nonlinear crystal for SPDC after undergoing frequency doubling. \cite{Rarity_2005}. In this configuration, three-fold coincidence counts are typically measured to plot the HOM dip. Visibility as high as $89\%$ has been achieved \cite{PhysRevA.83.031805}.\\
In those cases, it is important to ensure that the two photons arrive at the BS simultaneously. To achieve this, an optical path delay may need to be introduced into the photon paths within the experimental setup. This adjustment compensates for any difference in travel distance, thereby aligning the arrival times and optimizing the interference visibility.\\
Haldar et al. demonstrated two-photon interference between two independent SPDC sources, where each crystal was pumped by a continuous-wave laser \cite{Halder2007}. A key requirement for this interference is that the detection times must be measured with a precision finer than the coherence time of the photons. Since single-photon detectors inherently exhibit minimal jitter, the coherence time of the photons must be extended to surpass this jitter. This is achieved by applying narrow spectral filtering. In their experiment, a highly narrow bandpass filter with a bandwidth of $10$ picometers was used to ensure sufficient coherence.\\
Two-photon interference between independent single-photon sources has been demonstrated beyond SPDC. This includes interference between photons emitted from two vacancy centers within a nanodiamond \cite{PhysRevLett.113.113602,PhysRevLett.108.043604} or from two remotely positioned nanodiamonds \cite{WaltrichKlotzAgafonovKubanek+2023+3663+3669,PhysRevLett.108.143601}, as well as between two quantum dots \cite{PhysRevLett.104.137401}. Additionally, two-photon interference has been achieved using photons retrieved from two separate quantum memories \cite{Gera2024,PhysRevLett.129.093604}. These demonstrations highlight the versatility of HOM interference across various quantum systems.
\subsubsection{Entanglement-based communication}
Entanglement is a crucial element of quantum communication. Entanglement is generated by either producing two entangled particles from the same source or through direct interaction between particles. However, entanglement can also be established through a process known as entanglement swapping, which leverages the projection of two particles onto an entangled state. Notably, this approach does not necessitate any physical interaction between the particles involved. Instead, if each particle is already entangled with another partner, performing a measurement, such as a Bell-state measurement, on the partner particles can project the remaining two particles into an entangled state. This remarkable manifestation of the projection postulate forms the foundation of entanglement swapping \cite{PhysRevLett.80.3891,PhysRevLett.123.160501}.\\
A Bell-state measurement (BSM) is a process that projects two particles onto one of four maximally entangled states, known as Bell states. This measurement is essential for entanglement swapping as it determines the final entangled state of the remaining particles. The physical mechanism behind BSM often relies on two-photon interference, a phenomenon where two indistinguishable photons interfere destructively or constructively depending on their relative phase and polarization. When two photons arrive simultaneously at a beam splitter, the interference pattern generated dictates the projection onto a specific Bell state. By harnessing this interference, BSM can effectively entangle distant particles, making it a key ingredient in entanglement distribution protocols \cite{PhysRevA.53.R1209,PhysRevLett.71.4287}. BSM plays a vital role in quantum teleportation \cite{Bouwmeester1997}\\
Consider two single-photon sources. The first source generates an entangled photon pair, where both photons are either horizontal (HH) or vertical (VV) polarized, denoted as photons $1$ and $2$ with the state $\ket{\psi_1}=\frac{1}{\sqrt{2}}(\ket{HH}_{12}+\ket{VV}_{12}$. 
\begin{figure}[H]
\centering
\includegraphics[scale=0.5]{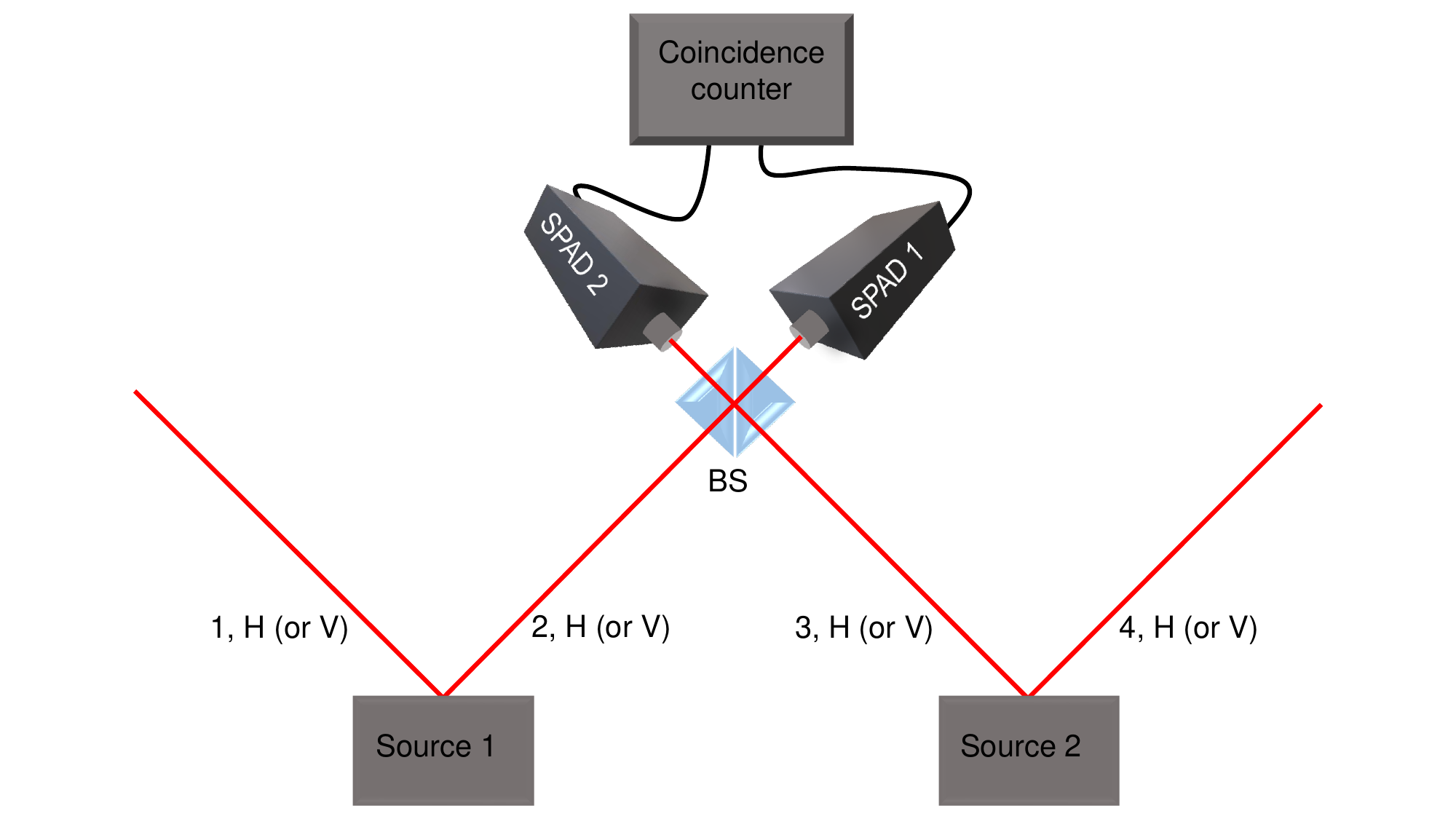}
\caption{The schematic illustrates the experimental setup for BSM.}
\label{fig4}
\end{figure}
The second source generates a similar entangled photon pair, also either HH or VV polarized, denoted as photons $3$ and $4$ with the state $\ket{\psi_1}=\frac{1}{\sqrt{2}}(\ket{HH}_{34}+\ket{VV}_{34}$. Thus, the total state of the system (photons $1$, $2$, $3$, and $4$) is the combined state of the two entangled pairs:
\begin{equation}
    \label{eq13}
\ket{\psi_{total}}=\ket{\psi_1}\otimes\ket{\psi_2}=\frac{1}{2}(\ket{HH}_{12}+\ket{VV}_{12}) \otimes (\ket{HH}_{34}+\ket{VV}_{34})
\end{equation}
Photons $2$ and $3$ undergo BSM, where they interfere on a balanced beam splitter. The BSM projects the state of photons $2$ and $3$ onto one of the four Bell states:
\begin{equation*}
    \ket{\phi^+}=\frac{1}{\sqrt{2}}(\ket{HH}+\ket{VV})
\end{equation*}
\begin{equation*}
    \ket{\phi^-}=\frac{1}{\sqrt{2}}(\ket{HH}-\ket{VV})
\end{equation*}
\begin{equation*}
    \ket{\psi^+}=\frac{1}{\sqrt{2}}(\ket{HV}+\ket{VH})
\end{equation*}
\begin{equation*}
    \ket{\psi^-}=\frac{1}{\sqrt{2}}(\ket{HV}-\ket{VH})
\end{equation*}
Now, considering the total state of photons $1$, $2$, $3$, and $4$, we can rewrite it in terms of the Bell states.
If photons $2$ and $3$ are projected into the $\ket{\phi^+}$ state, the total state collapses to:
\begin{equation}
    \label{eq14}
    \ket{\psi_{total}}=\frac{1}{2}(\ket{\phi^+}_{23})\otimes(\ket{HH}_{14}+\ket{VV}_{14})
\end{equation}
This process results in photons $1$ and $4$ becoming entangled. The final entangled state between photons $1$ and $4$ is, $\ket{\psi}_{14}=\frac{1}{\sqrt{2}}(\ket{HH}_{14}+\ket{VV}_{14})$. If photons $2$ and $3$ are measured in the $\ket{\phi^-}$ state, the result is $\ket{\psi}_{14}=\frac{1}{\sqrt{2}}(\ket{HH}_{14}-\ket{VV}_{14})$. Therefore, the entanglement-swapping process depends on the BSM outcome.\\
Entanglement swapping plays a critical role in developing quantum repeater networks, which are essential for extending the range of quantum communication. In a quantum repeater architecture, quantum memories are distributed along a communication channel, with each memory holding two atoms capable of emitting photons in opposite directions. A Bell-state analyzer positioned midway between two adjacent memories detects the photons travelling from each memory, performing a BSM to establish entanglement between the atoms stored in the two memories. By repeating this process across successive links, entanglement can gradually be extended across multiple memories, effectively connecting distant locations. Once successful, this chain of entanglement enables the creation of entangled pairs between the first and final nodes of the network, facilitating secure quantum communication over long distances \cite{PhysRevA.71.050302,PhysRevA.92.012329,doi:10.1126/science.1221856}. A detailed and comprehensive exploration of entanglement swapping in quantum repeater network architecture would warrant a separate review paper. Readers are encouraged to refer to \cite{RevModPhys.84.777,RevModPhys.95.045006} for further information.
\subsubsection{Signature of precision}
As previously mentioned, the HOM dip serves as a key indicator of photon indistinguishability. It provides a means to quantify the optical delay between two photons. The width of the HOM dip is inherently broad, and this width can be precisely controlled, making it particularly effective for detecting and measuring small optical delays. The control over the HOM dip width allows for fine-tuning of the system to improve precision in timing measurements. In the case of SPDC, the width of the HOM dip can be adjusted by varying the crystal length \cite{Singh:21}. Additionally, the bandwidth of the bandpass filter can also be adjusted to control the HOM dip width \cite{doi:10.1126/sciadv.aap9416}. In a common-path HOM setup described in \cite{PhysRevA.62.063808}, researchers successfully measured the group delay of photons with an impressive precision of $0.1$ femtoseconds and the phase delay with a precision of $8$ attoseconds. Further advancements in precision measurement were introduced by Lyons et al. in their work on attosecond-resolution HOM interferometry \cite{doi:10.1126/sciadv.aap9416}. They devised a measurement strategy based on Fisher information analysis, which enabled the team to achieve an accuracy of $0.5$ attoseconds.\\
The HOM dip can be effectively applied to measure and control the group velocity of photons by manipulating their transverse profiles. By structuring the signal photon’s wavefront using spatial light modulators, it is possible to induce changes in the photon’s group velocity. This is achieved by altering the transverse spatial characteristics of the photon, such as its wavefront shape.
After the photon travels through free space, a second Spatial light modulator is used to reverse the wavefront modification, thereby restoring the photon to its original state. The interference between the structured signal photon and the idler photon, when both are directed onto a balanced BS, produces the HOM dip. Importantly, when the group velocity of the signal photon is changed, the HOM dip shifts. This shift in the HOM dip directly corresponds to a time delay, providing a clear measurement of the change in the photon’s group velocity \cite{doi:10.1126/science.aaa3035}. An experiment described in \cite{Lyons:18} further demonstrates this concept. The researchers showed that a change in the orbital angular momentum of a photon introduces a time shift in the HOM dip. By modifying the orbital angular momentum of the signal photon, the HOM dip undergoes a shift, effectively linking the photon’s orbital angular momentum to a temporal delay. This highlights the potential of HOM interference as a powerful tool for measuring and controlling photon velocities. \cite{PhysRevLett.132.193603} identifies an optimal scaling relationship between the precision and visibility of the HOM dip. Therefore, HOM interference can be used in quantum metrology to achieve measurements with exceptionally high precision, thus opening a new regime in precision measurement techniques. However, the detailed discussion on this topic, including its applications and theoretical framework, is outside the scope of this review paper. For further reading, readers are advised to refer to \cite{Chen2022-nc}.\\
Optical Coherence Tomography (OCT) is an imaging technique used for depth profiling of tissue samples. It works by measuring the interference between light reflected from the sample and a reference beam. In practice, this technique is limited by the coherence of the light source and the dispersion of the sample. In this context, HOM interference offers a significant improvement. To overcome the limitations, a technique called Quantum Optical Coherence Tomography (QOCT) has been proposed. This technique leverages the automatic dispersion cancellation inherent in the two-photon interference effect \cite{PhysRevA.65.053817}. Additionally, QOCT offers a twofold improvement in resolution compared to classical OCT. In the experimental setup, the sample is placed in one arm of the HOM interferometer, and coincidence counts are recorded while varying the path of the other arm using a translation stage. Shortly after its proposal, QOCT was experimentally demonstrated, achieving a fivefold improvement in resolution compared to OCT \cite{PhysRevLett.91.083601}. Various experiments using different techniques have achieved even better depth resolution. However, a detailed discussion of these experiments is beyond the scope of this review. For further information, readers are advised to consult \cite{JIN2024100519}.

\section{Multiple photon path interference}
\subsection{Interference and sum rule}
In his seminal paper \cite{doi:10.1142/S021773239400294X} Rafael D. Sorkin examines how interference emerges as a consequence of the failure of classical probability additivity. It introduces a hierarchy of sum rules, where the second sum rule corresponds to classical probability and ensures no interference. However, quantum probabilities obey a weaker condition where interference terms persist. The second sum rule refers to the classical additivity of probabilities:
\begin{equation}
    \label{higher_order_eq1}
    P(A\cup B)= P(A)+P(B)
\end{equation}
for mutually exclusive events $A$ and $B$. In classical probability theory, this always holds, meaning there is no interference between different paths or events. However, in quantum mechanics, probabilities do not always sum linearly due to interference effects. The deviation from this classical additivity is captured by an interference term:
\begin{equation}
    \label{higher_order_eq2}
    I(A,B)=P(A\cup B)-P(A)-P(B)
\end{equation}
To derive the third-order sum rule for a triple-slit experiment, the Born rule and the superposition principle are used. According to the Born rule, the probability of an event is given by the squared modulus of the wavefunction: $P(X)=\vert \psi(X)\vert^2$. For a particle passing through slits $A$, $B$, and $C$, the total wavefunction is the sum of contributions from each slit: $\psi_{A,B,C}=\psi_A+\psi_B+\psi_C$. For two slits $A$ and $B$, the probability is: $P(A\cup B)=\vert \psi_A+\psi_B\vert^2$. Expanding using the Born's rule, the above equation becomes,
\begin{equation}
    \label{higher_order_eq3}
 P(A\cup B)=\vert \psi_A\vert^2+\vert \psi_B\vert^2+2 \mathcal{R}(\psi_A^*\psi_B)   
\end{equation}
Here, $\mathcal{R}(\psi_A^*\psi_B)$ is the interference term between slits $A$ and $B$. Similarly, for other pairs:
\begin{equation}
\label{higher_order_eq4}
 P(B\cup C)=\vert \psi_B\vert^2+\vert \psi_C\vert^2+2 \mathcal{R}(\psi_B^*\psi_C)
 \end{equation}
 \begin{equation}
 \label{higher_order_eq5}
  P(A\cup C)=\vert \psi_A\vert^2+\vert \psi_C\vert^2+2 \mathcal{R}(\psi_A^*\psi_C)   
 \end{equation}
For all three slits open: 
\begin{equation}
\label{higher_order_eq6}
P(A 
 \cup B \cup C)=\vert \psi_A+\psi_B+\psi_C\vert^2=\vert\psi_A\vert^2+\vert \psi_B\vert^2+\vert \psi_C\vert^2+2 \mathcal{R}(\psi_A^*\psi_B)+2 \mathcal{R}(\psi_B^*\psi_C)+2 \mathcal{R}(\psi_A^*\psi_C)
\end{equation}
The third-order interference term is defined as:
 \begin{equation}
     \label{higher_order_eq7}
     I(A,B,C)=P(A \cup B \cup C)-P(A \cup B)-P(B \cup C)-P(A \cup C)+P(A) + P(B) +P(C)
 \end{equation}
 Substituting Eqs. \ref{higher_order_eq6},\ref{higher_order_eq5},\ref{higher_order_eq4},and \ref{higher_order_eq3} in Eq. \ref{higher_order_eq7}, and expanding all terms, $I(A,B,C)=0$.
 Therefore, $I(A,B,C)=0$ serves as a test for the validity of the Born rule. In this context, the Sorkin parameter quantifies any deviation from the expected quantum probability sum, measuring the extent to which the observed probability differs from the classical sum of individual slit probabilities.\\
 In \cite{doi:10.1126/science.1190545}, the authors have conducted a three-slit experiment with photons to test the validity of Born’s rule by investigating the presence of third-order interference. They measured the quantity $\epsilon$, defined as
 \begin{equation}
     \label{higher_order_eq8}
     \epsilon=P_{ABC}-(P_{AB}+P_{AC}+P_{BC})+(P_A+P_B+P_C)-P_0
 \end{equation}
 where $P_{ABC}$ is the probability of detection when all three slits are open, $P_{AB}$, $P_{AC}$, $P_{BC}$ are probabilities when two slits are open, $P_A$, $P_B$, $P_C$ correspond to single-slit openings, and $P_0$ accounts for background noise (probability of detection when all slits are closed). To compare any possible deviation from the rule, the study defines a normalized quantity: $\kappa=\frac{\epsilon}{\delta}$, where $\delta$ is defined as,
 \begin{equation*}
 \begin{split}
     \delta &=\vert I_{AB}\vert+\vert I_{AB}\vert+\vert I_{AB}\vert\\
     &= \vert P_{AB}-P_A-P_B+P_0\vert+\vert P_{BC}-P_B-P_C+P_0\vert+\vert P_{AC}-P_A-P_C+P_0\vert\\
 \end{split}
 \end{equation*}
If $\kappa=0$, Born’s rule holds perfectly, whereas $\kappa \ne 0$  suggests a deviation from it. The experiment employed an optical setup with a triple-slit, where each slit could be selectively opened or blocked using a blocking mask, with slit combinations chosen randomly to minimize systematic errors. Various types of photon sources were used, and the interference patterns were recorded using SPADs. By analyzing photon intensities across all slit configurations (single-, double-, and triple-slit cases), the researchers established an upper bound on three-path interference, limiting it to less than $10^-2$ of the expected two-path interference for SPDC based single photons and constraining it to $10^-3$ for the attenuated coherent laser beam. The results were consistent with Born’s rule, effectively ruling out higher-order interference with high precision.\\
Several experiments have since been conducted to test Born’s rule.  In \cite{Sollner2012} the authors present a different approach using a three-path interferometer to establish a tighter empirical upper bound on potential deviations. Unlike conventional slit-based setups, this experiment employs a transmission grating to generate three independent optical paths. Two paths pass through individually controlled phase plates, allowing precise phase adjustments. All three paths are then retroreflected using a common mirror. To separate the outgoing and incoming beams, the setup incorporates a double pass through a quarter-wave plate and a polarizing BS. To account for detector nonlinearity, the researchers model the response using a Poissonian light source and a nonlinearity equation, allowing them to predict and subtract systematic errors. By measuring and comparing the three-path interference pattern with theoretical predictions, they set a new upper bound on three-path interference: $< 0.0015 \pm 0.0029$. Improved precision was achieved \cite{Kauten_2017} by using a stabilized five-path interferometer instead of a three-path setup, minimizing systematic errors such as detector nonlinearity and phase drift. Two diffractive BS were employed to create and recombine five independent paths, with shutter assemblies providing control over which paths were open. Phase plates, mounted on motorized rotation stages, were inserted in each path for precise phase control. Detector nonlinearity was modelled and calibrated by separately measuring response curves. The entire interferometer was enclosed and temperature-stabilized. This study set tighter upper bounds on higher-order interference at: $10^-3$. $\kappa=0$ has also been measured in various systems beyond photonics, including matter wave, atomic tests, and nuclear magnetic resonance systems. However, these are beyond the scope of this review article.
\subsection{Non-zero Sorkin parameter}
\begin{figure}
\centering
\includegraphics[scale=0.6]{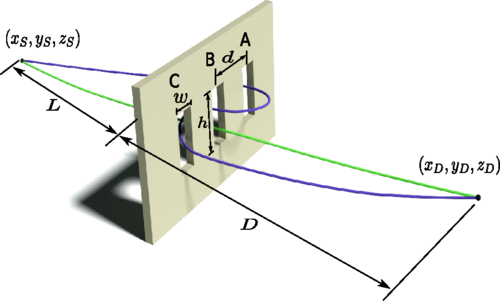}
\caption{A point source is located at $(x_s,y_s,z_s)$ at a distance $L$ from a triple-slit system. Each slit has a width $w$, height $h$, and an inter-slit separation $d$. The detection point is at $(x_d,y_d,z_d)$ situated at a distance $D$ from the slit. The green path represents the classical trajectory, while the purple loops around the slits illustrate non-classical paths \cite{PhysRevLett.113.120406}.}
\label{higher_order_fig}
\end{figure}

A nonzero $\epsilon$ signals a departure from standard quantum mechanics, indicating the presence of third-order interference. However, a nonzero $\epsilon$
does not necessarily imply a violation of Born's rule. The double-slit experiment, a cornerstone of both optics and quantum mechanics, typically assumes the wave function with both slits open is the sum of the individual wave functions from each slit: $\psi_{AB}=\psi_A+\psi_B$.  However, this assumption is mathematically inaccurate. The three scenarios mentioned above correspond to distinct boundary conditions, meaning that the superposition principle can only be applied as an approximation in these cases. This can be addressed by Feynman’s path integral formalism. This formalism involves summing over all possible paths a particle can take through the two slits. It includes not only the near-straight trajectories from the source to the detector through either slit (green paths in Fig. \ref{higher_order_fig}), but also incorporates nonclassical paths, such as the looped ones shown in purple in Fig. \ref{higher_order_fig}. While these looped paths contribute much less to the total intensity at the detector compared to the straight-line paths, their effect is nonzero and finite \cite{PhysRevLett.113.120406}.
These contributions lead to a modified wave function $\psi_{AB}=\psi_A+\psi_B+\psi_L$, where $\psi_L$ represents the looped paths. The presence of nonclassical paths, as shown in Fig. \ref{higher_order_fig}, challenges the conventional application of the superposition principle in interference experiments.
\cite{PhysRevLett.113.120406} suggests that the triple-slit experiment can be used to quantify the impact of nonclassical paths. These paths also affect the calculation of the experimentally measurable quantity $\kappa$, which has been used to set bounds on higher-order interference terms in quantum mechanics. $\kappa$ exhibits modulation in its nonzero values as a function of the detector position. This means that if one plots the triple-slit interference pattern—intensity versus position— $\kappa$
varies along the spatial coordinate.
If only classical paths contributed to interference, $\kappa$ would be zero. However, considering nonclassical paths allows for nonzero values of $\kappa$, indicating that a nonzero $\kappa$ does not falsify Born’s rule but instead supports the broader Feynman path integral formalism. The experiment in Ref.\cite{doi:10.1126/science.1190545} did not detect the expected nonzero value of $\kappa$ due to systematic errors. However, these errors can be corrected in future experiments, enabling more accurate measurements. Further analysis reveals that $\kappa$ is highly sensitive to experimental parameters, and there is potential for larger $\kappa$ values under certain conditions, such as increasing the wavelength of the photons.\\
An analytic formula has been derived to establish a bound on $\kappa$ as a function of detector position in the far-field diffraction regime \cite{Sinha2015}. This formula has been verified against results from Finite Difference Time Domain simulations and numerical integration, demonstrating close agreement. The derived bound is: $\vert \kappa_{max}\vert \approx 0.03 \frac{\lambda^{\frac{3}{2}}}{d^{\frac{1}{2}}w}$, where $\lambda$ is the wavelength of the light or particle used in the experiment.\\
The experimental verification of looped trajectories in a triple-slit interference experiment was achieved by enhancing electromagnetic near-fields around the slits using surface plasmons \cite{Magana-Loaiza2016}. A nanostructured gold film with three slits in different opening configurations was fabricated to support surface plasmon excitation. A heralded single-photon source was used, and photon polarization was controlled with a half-wave plate and a polarizer to either excite surface plasmons (x-polarization) or suppress them (y-polarization). For x-polarization, the electric field oscillates parallel to the plane of the slits, inducing electron motion along the metal film and exciting surface plasmon waves. These plasmons enhances near-field interactions, increasing coupling between slits and amplifying the probability of looped trajectories. In contrast, for y-polarization, where the electric field oscillates perpendicular to the slit plane, surface plasmons are not excited, and only direct (classical) paths contributes, making looped trajectories negligible. As a result, the y-polarized interference pattern follows a standard three-slit distribution with no visible looped paths, while the x-polarized pattern shows significant deviations, with increased visibility due to interference between classical and looped trajectories. The probability ratios for x- and y-polarized photons varies across different slit configurations, confirming that looped paths influences the pattern differently under varying conditions. The measured $\kappa$ remained close to zero for y-polarization but became significantly nonzero for x-polarization, demonstrating that looped trajectories contribute to interference fringes when near-field effects are enhanced through material-induced enhancement.
\begin{figure}[H]
\setlength{\abovecaptionskip}{10pt} 
\centering
\includegraphics[scale=1.0]{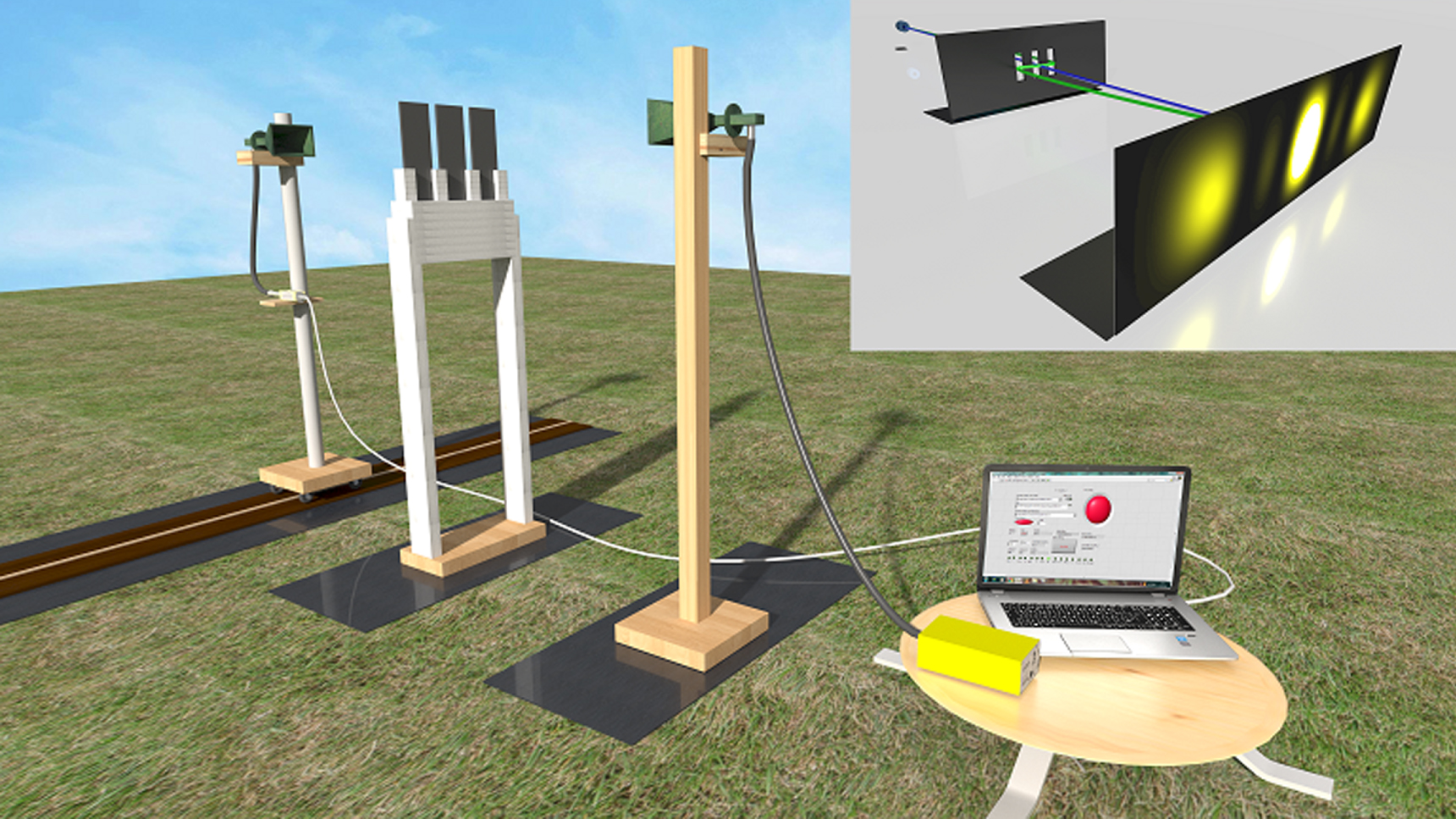}
\caption{Schematic of the experimental setup in \cite{Rengaraj_2018}. The green antennas on both sides are pyramidal horn antennas, serving as the source and receiver of electromagnetic waves at $5$ cm wavelength. The receiving antenna is mounted on a movable rail, allowing for the measurement of diffraction patterns. Positioned between the source and receiver are three slots. The inset illustrates a triple-slit schematic, where the blue line represents a classically dominant path, while the green line depicts a sub-leading path within the path integral formalism.}
\label{higher_order_fig2}
\end{figure}
While the $\kappa$ was enhanced by utilizing near-field components of the photon wave function and material-induced effects in \cite{Magana-Loaiza2016}, a non-zero $\kappa$, purely attributed to length scale-dependent boundary condition effects on the superposition principle, was observed in a completely different wavelength domain in \cite{Rengaraj_2018}. The experiment was designed to control the effects of non-classical paths, allowing the researchers to increase or decrease the effect by adding or removing obstructions, providing definitive proof of their existence. A precision triple-slot setup as shown in Fig.\ref{higher_order_fig2} in the centimeter-wave domain was employed, using pyramidal horn antennas and specially designed composite materials as microwave absorbers. To minimize electromagnetic noise and interference, the experiment was conducted in a remote location, effectively simulating an anechoic chamber. Since etching slits in an absorbing layer at this scale was impractical, the experiment used absorbing slots surrounded by free space to approximate infinitely large boundaries. 
The modified expression for $\kappa$ is defined based on the magnitude of the Poynting vector at specific detector positions and is calculated by measuring power values for different slot combinations,
\begin{equation}
    \label{higher_order_eq9}
    \kappa'=\frac{P_0-(P_{ABC}-P_{AB}-P_{BC}-P_{AC}+P_A+P_B+P_C)}{max(P_0)}
\end{equation}
A pyramidal horn antenna, operating at a $5$ cm wavelength, emitted electromagnetic waves directed at $10$ cm-wide slots with a $13$ cm inter-slot distance. The slots were constructed from composite materials: two layers of Eccosorb SF6.0 (a near-perfect microwave absorber) with an aluminium layer sandwiched in between to enhance absorption, particularly of back-reflected beams. A detector horn antenna, mounted on a moving rail, collected power measurements at various positions. The source and detector were placed $2.5$ meters apart ($1.25$ meters between the source and slot plane, and the same distance between the slot and detector plane), and the values of $\kappa'$ showed modulation as a function of detector position.To ensure that the measurement time scale was sufficient, an antenna radiation pattern was recorded before measuring each slot combination. The background patterns remained stable throughout, confirming that noise fluctuations occurred on a much longer timescale than the measurement period. Background levels were averaged before and after each measurement to minimize bias, and the order of slot combinations was randomized to eliminate systematic errors. To independently verify that the observed $\kappa'$ was not due to unaccounted experimental errors, absorbers (baffles) were introduced perpendicular to the slots to gradually suppress the effects of non-classical paths. As suggested in \cite{PhysRevLett.113.120406}, "hugging paths"—which cross the slit plane twice—are the primary contributors to $\kappa'$. By systematically increasing the baffle size, a clear decrease in $\kappa'$ was observed, confirming that hugging paths dominate its non-zero value.\\
A key potential error in observing a non-zero $\kappa'$ is detector non-linearity—if the detector response deviates from linearity with increasing power, it could introduce a false non-zero. To rule this out, a detailed analysis using spline interpolation and polynomial fitting was conducted. The non-linearity effects estimated through these models were found to be at least two to three orders of magnitude smaller than the measured $\kappa'$, demonstrating that detector non-linearity could not account for the observed values. Additionally, potential non-linearity in the source was ruled out, as the source was operated at a constant power.\\
While the slit experiment typically results in a small third-order interference term, it was theoretically shown \cite{PhysRevA.103.052204} that higher-order interference can arise within the standard second quantization framework due to nonlinear effects in multipartite interactions. An increased value of nonzero third-order interference can be observed with current technologies, such as nonlinear optics or Bose-Einstein condensates. \cite{PhysRevA.107.032211} demonstrated an experiment using a `nonlinear triple slit', which consists of three laser beams interacting in an optically nonlinear crystal. The authors showed that higher-order interference can be turned on and off by modulating the nonlinear interactions in the system. A beam with a wavelength of nanometers is used, and the parameter $\kappa$ is found to be $0.0334\pm 0.0002$.A study \cite{PhysRevResearch.2.012051} has established that interference of order $(2M+1)$ and higher vanishes for $M$ particles. Furthermore, it introduced generalized many-particle Sorkin parameters, which are predicted to be zero if Born’s rule holds. Many-particle higher-order interference has been experimentally demonstrated using a five-slit setup with a coherent state at a $633$ nm wavelength, where the mean photon number was varied \cite{PhysRevLett.126.190401}. The study demonstrated that fifth-order interference is constrained to $10^-3$ in the intensity-correlation regime and $10^-2$ in the photon-correlation regime. The search for a genuine non-zero Sorkin parameter continues, pushing the boundaries of our understanding of higher-order interference. Future experiments with advanced photon sources and detection techniques may provide deeper insights into this fundamental question.

\section{Conclusion and Outlook}
The exploration of interference phenomena stands as a cornerstone in our understanding of quantum mechanics. From the fundamental second-order correlation measurements ($g^{(2)}$) to the more sophisticated Hong-Ou-Mandel interference effects, these experimental observations not only validate theoretical frameworks but also challenge our classical intuitions about the nature of reality.
The $g^{(2)}$ correlation function has proven invaluable for distinguishing quantum from classical light, providing a robust metric for characterizing non-classical photon statistics. Meanwhile, the HOM effect—where identical photons entering a BS through different ports demonstrate fourth-order interference by emerging together—has become emblematic of quantum behavior with no classical analog. Yet, as discussed in this article, even this quintessentially quantum phenomenon can be mimicked by classical fields through careful phase control between classical pulses, blurring the boundary between quantum and classical descriptions.
This remarkable ability to reproduce ostensibly quantum effects through classical means raises profound questions about the true nature of the quantum-classical divide. Rather than viewing these domains as fundamentally separate, perhaps we should understand them as different mathematical frameworks describing the same underlying reality, with quantum mechanics offering a more general description that reduces to classical physics under appropriate conditions.
The investigation of higher-order interference, initiated by Sorkin's hierarchical framework, represents the next frontier in this ongoing inquiry. Sorkin's parameter $\kappa$, designed to quantify departures from the quantum superposition principle, offers a tantalizing window into possible physics beyond standard quantum mechanics. Despite numerous experimental efforts, a conclusive measurement of a non-zero Sorkin parameter that withstands all experimental scrutiny remains elusive. The technical challenges are formidable—requiring unprecedented precision in phase stability, detector efficiency, and background suppression—yet the potential rewards are equally substantial.
Should future experiments definitively establish a non-zero Sorkin parameter, the implications would be revolutionary, potentially necessitating modifications to quantum theory itself. Conversely, increasingly precise null results would further cement quantum mechanics as a complete description of interference phenomena, placing ever-tighter constraints on alternative theories.
As experimental techniques continue to advance, particularly in integrated photonics, quantum dot single-photon sources, and superconducting detector technology, we stand at the threshold of resolving these fundamental questions. The story of interference in quantum mechanics thus remains unfinished—a vibrant field of inquiry where fundamental questions about the nature of reality continue to drive both theoretical innovation and experimental ingenuity.
The path forward lies not merely in refining existing experiments, but in developing entirely new paradigms for probing quantum interference at ever-higher orders and with unprecedented precision. Whether these investigations ultimately reveal physics beyond standard quantum mechanics or further validate its remarkable descriptive power, they will undoubtedly deepen our understanding of the fundamental principles governing our universe.

\section*{Funding}
US acknowledges partial support provided by the National Quantum Mission of the Department of Science and Technology.
\section*{Data availability}
This manuscript has no associated data, as it is a theoretical review.
\section*{Conflict of interest}
The authors declare that there are no conflicts of interest or competing interests directly or indirectly related to this work.
\bibliographystyle{unsrt}
\bibliography{ref}

\end{document}